\documentclass{aa}

\usepackage{graphicx}
\usepackage{times}

\begin{document}

   \thesaurus{06         
              (13.07.1)}  
            \title{Archival search for recurrent activity at the position
              of the gamma-ray burst GRB 970228 optical counterpart.}



   \author{J. Gorosabel
          \inst{}
   \and A. J. Castro-Tirado
          \inst{}
          }

   \offprints{J. Gorosabel (jgu@laeff.esa.es)}

   \institute{Laboratorio de Astrof\'{\i}sica Espacial y F\'{\i}sica 
              Fundamental (LAEFF-INTA), P.O. Box 50727,
              E-28080 Madrid, Spain}

   \date{Received date; accepted date}

  \titlerunning{Archival searches for GRB 970228.}

  \authorrunning{Gorosabel \& Castro-Tirado}

   \maketitle

   \begin{abstract}
     We have examined 8004 plates at the Harvard College Observatory Plate
     Collection searching for optical transient emission from the gamma-ray
     burst GRB 970228.  This is the first archival search carried out so
     far for a gamma-ray burst with known transient optical emission.  The
     total exposure time amounts to $\sim 1.1$ yr. No convincing optical
     activity was found above $12.5$ mag at the expected position of the
     GRB 970228 optical counterpart.

   \end{abstract}

%

\section{INTRODUCTION}

GRB 970228 was detected as a 5.5-s brief, high energy event on 28 February
1997, by the satellite BeppoSAX (Costa et al. 1997a). A decaying X-ray
emission (the X-ray afterglow) was found 8 hr after the onset of the event
(Costa et al. 1997b). Optical observations started 15.4 hr after the burst
(Pedichini et al. 1997), which allowed the search of variable objects in
the intersection between the $\gamma$-ray error box, the X-ray error box
and the Interplanetary Network (IPN) annulus (Hurley et al.  1997).  They
led to the discovery of the first optical counterpart of a GRB (van
Paradijs et al.  1997).

The recent measurement of the redshift of the GRB 970508 optical
counterpart has established that most GRBs - if not all - lie at
cosmological distances (Metzger et al.  1997). While galactic GRB models
allow multiple outbursts from each source (Li and Dermer 1992), the
cosmological models are usually based on singular, cataclysmic highly
energetic events.  Thus, the detection of recurrent optical emission for a
GRB would impose severe constraints on the cosmological models (Narayan,
Paczy\'nki and Piran 1992, M\'esz\'aros and Rees 1993).  Archival searches
can be crucial in order to clarify this problem.

Archival searches in GRB error boxes rely upon the assumption that GRBs do
repeat and show optical transient emission. With perhaps the exception of
OT 1928 (Schaefer 1981, Hudec et al. 1994), no firm candidates have yet
been established (Hudec et al. 1993, and references therein, also Gorosabel
and Castro-Tirado 1998). The detection of the two first optical
counterparts for GRB 970228 (van Paradijs et al. 1997, Guarnieri et al.
1997) and GRB 970508 (Bond 1997,Castro-Tirado et al. 1998) has been an
important breakthrough.  However, for some other GRBs no optical emission
was detected although very fast and deep follow up observations were
carried out (Castro-Tirado et al.  1997, Groot et al. 1998).

\section{THE ARCHIVAL SEARCH}

In order to search for optical recurrent activity at the position of the
GRB 970228 optical counterpart we have examined 8004 plates at the Harvard
College Observatory Plate Collection (HCO). Table~1 displays the list of
plates examined at HCO.  The plates span 64 yr, from 1889 to 1952, and the
total exposure time amounts to $\sim 1.1$ yr. The accurate position of the
optical counterpart made the archival search very easy in comparison to
other GRBs with larger error boxes.

Many spots were found on the plates, but none of them was consistent with
the position of the true optical counterpart, except for a $\sim$ 9 mag
star-like spot found on the plate RH 4888 (see Fig. 1). This plate was
taken on 18 March 1933 and represents three exposures of 6 min each.  Two
of the three exposures included the GRB 970228 field and are shifted $\sim
4^{\prime}.8$ from each other. The third one includes the north equatorial
pole.  As expected for a real optical transient, we found two spots
corresponding -within the astrometrical errors- to both the real and
"shifted" positions of the true GRB 970228 optical counterpart.  However,
an accurate measurement of the two positions revealed that they are not
trailed in a similar way to other stellar images. We then realized that the
two spots correspond to a pair of stars near the north pole that were
imaged during the third exposure.

\begin{table}[bt]
\caption{Plates examined at HCO for GRB 970228}
\begin{center}
  \begin{tabular}{cccccc}
\hline
{\scriptsize Plate}&{\scriptsize Limiting}&{\scriptsize Dates}&{\scriptsize Number}&{\scriptsize Total exposure}&{\scriptsize Limiting}\\ 
{\scriptsize series}&{\scriptsize mag.}&&{\scriptsize of
  plates}&{\scriptsize time (hr)}&{\scriptsize mag. (1-s burst)}\\ 
\hline
{\scriptsize A        }&{\scriptsize  16.5 }&{\scriptsize 1893-1950}&{\scriptsize   3   }&{\scriptsize 0.5 }&{\scriptsize  8.2 }\\
{\scriptsize AC       }&{\scriptsize  14.1 }&{\scriptsize 1898-1957}&{\scriptsize  2906 }&{\scriptsize 3834 }&{\scriptsize  5.8 }\\
{\scriptsize AI,BI,FA }&{\scriptsize  12.5 }&{\scriptsize 1900-1953}&{\scriptsize  4602}&{\scriptsize 5287}&{\scriptsize  4.2 }\\
{\scriptsize MA       }&{\scriptsize  16.5 }&{\scriptsize 1905-1952}&{\scriptsize   5   }&{\scriptsize 5.2 }&{\scriptsize  9.0 }\\
{\scriptsize MC       }&{\scriptsize  16.5 }&{\scriptsize 1909-1952}&{\scriptsize   7   }&{\scriptsize 5.4 }&{\scriptsize  9.0 }\\
{\scriptsize MF       }&{\scriptsize  16.4 }&{\scriptsize 1915-1952}&{\scriptsize  35   }&{\scriptsize 14.2 }&{\scriptsize  8.9 }\\
{\scriptsize RH       }&{\scriptsize  14.8 }&{\scriptsize 1928-1952}&{\scriptsize  446  }&{\scriptsize 472 }&{\scriptsize  6.5 }\\

\hline

\end{tabular} 
\end{center}
\end{table}

In order to estimate the rate of plate faults, we investigated the number
of spots found on the AC plates in a $15^{\prime}$ radius region around the
OT position (see Fig.~2).  The rate of plate faults with 0.2--0.6 mag
above the background is $\sim 8 \times 10^{-4}$ mm$^{-2}$ plate$^{-1}$, a
value comparable to that reported by Greiner et al.  (1987).
\DeclareGraphicsRule{.jpg}{eps}{.jpg.bb}{`convert #1 'eps:-'}
 \begin{figure}[t]
   \centering
   \fbox{\includegraphics[totalheight=11.0cm,width=6.5cm]{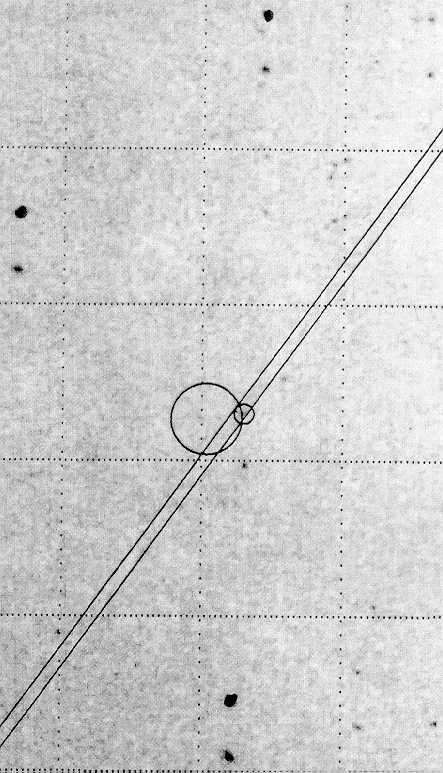}}
    \caption{The region of the HCO plate RH 4888 that includes the position of 
      the GRB 970228 optical counterpart (at the center of the image). Both
      stars seen in the third exposure (see detailed explanation in the
      main text) coincide with the position of the true counterpart, but
      only one of them is shown here within the intersection of the IPN and
      WFC error boxes. North is at the top and east to the left. The field
      of view is $\sim 38^{\prime} \times 66^{\prime}.$}
 \end{figure}  
\section{DISCUSSION AND CONCLUSIONS}

 The $\sim$ 8000 plates span $\sim$1.1 yr.  Many star-like spots were
 found.  With the exception of the above-mentioned case, none of them was
 consistent with the position of~~the~~true GRB 970228 optical counterpart.
 Therefore we can settle a lower limit of $\sim$ 1.1 yr for any recurrent
 optical transient emission activity brighter than 12.5 mag (or 4.2 mag if
 a 1-s flash is assumed). According to extrapolation of the gamma-ray
 spectra into longer wavelengths for the strongest bursts observed by
 BATSE, the magnitude of the optical flash that -eventually -would arise
 simultaneously to the gamma-ray event, could reach a magnitude of 2.5 in
 1-s (for the spectral index $\alpha$ $\geq$ --1, Ford and Band, 1996).  If
 this would have been the case for a previous bursting activity at the GRB
 970228 location, such an optical flash would have been easily detected in
 any of the HCO plates.

\section{Acknowledgements}
The authors wish to thank R. Hudec for fruitful conversations. We are also
indebted to the HCO plate curator M. L. Hazen, and to A. Doane for the
facilities given at the HCO plate stacks.
 \begin{figure}[h]
   \centering
   \fbox{\includegraphics[totalheight=7.5cm,width=5.0cm,angle=-90]{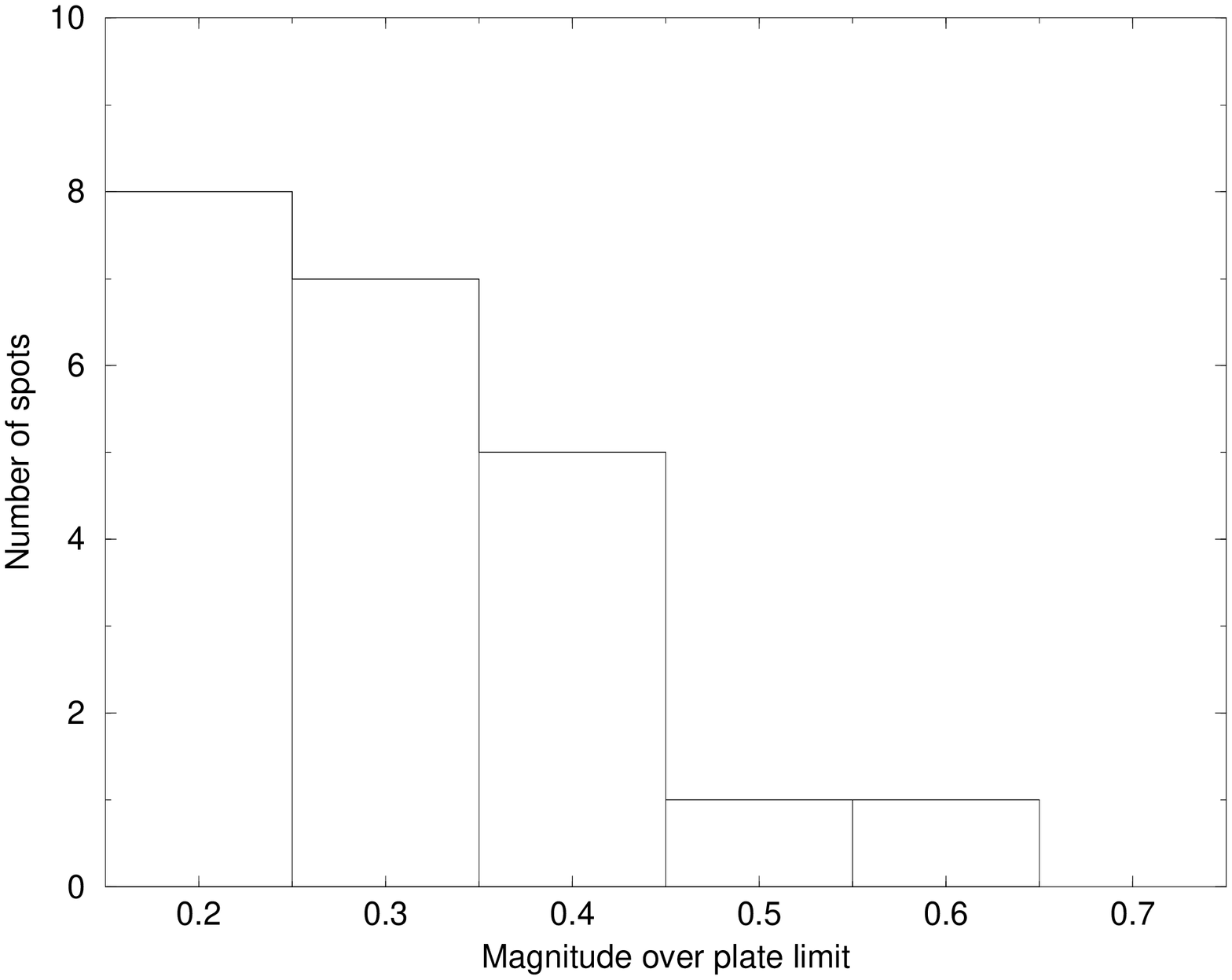}}
    \caption{Histogram of the 22 plate faults found in a $\sim 15^{\prime}$
      radius region around the OT position on the AC plate series.}
 \end{figure}


\begin{thebibliography}{}
\bibitem{Bond97} Bond, H., 1997, IAU Circ. 6654.

\bibitem{Cas97} Castro-Tirado, A.J., et al., 1997, IAU Circ. 6598.

\bibitem{Cas98} Castro-Tirado, A.J., et al., 1998, Sci 279, 1011.

\bibitem{Costa97a} Costa, E., et al., 1997a, IAU Circ. 6572.

\bibitem{Costa97b} Costa, E., et al., 1997b, IAU Circ. 6576.

\bibitem{Ford96} Ford, L.A., \& Band, D.L., 1996, ApJ 473, 1013.

\bibitem{Goro98} Gorosabel, J., \& Castro-Tirado, A.J., 1998, A\&A,
  in press.

\bibitem{Grein87} Greiner, J., Flohrer, J., Wenzel, W., \& Lehmann, T.,
  1987, Astroph. Space. Sci. 138, 155.

\bibitem{Groot98} Groot, P.J., et al., 1998, ApJL, in press.

\bibitem{Gua97} Guarnieri, A., et al., 1997, A\&A 328, L13.

\bibitem{Hude93} Hudec, R., 1993, A\&AS 97, 49.

\bibitem{Hude94} Hudec, R., Pravec, P., \& Borovi\u{c}ka J., 1994, A\&A
  284, 499.

\bibitem{Hur97} Hurley, K., et al., 1997, ApJ 485, L1.

\bibitem{Li92} Li, H, \& Dermer, C.,1992, Nat. 259, 514.

\bibitem{Mesz93} M\'esz\'aros, P., \& Rees, M.J., 1993, ApJ 405, 278.

\bibitem{Metz97} Metzger, et al., 1997, Nat. 387, 878.

\bibitem{Nara92} Narayan, R., Paczy\'nski, B., Piran, T. 1992, ApJ 395, L83

\bibitem{Ped97} Pedichini, F., et al., 1997, A\&A 327, L36.

\bibitem{Scha81} Schaefer, B.E., 1981, Nat. 294, 722.

\bibitem{van97} van Paradijs J., et al., 1997, Nat. 386, 686.

\end{thebibliography}
\end{document}